 \documentclass[aps, oneocolumn, showpacs, amsmath]{revtex4-2}
\usepackage{dcolumn}
\usepackage{bm}
\usepackage{graphicx}
\usepackage{color}
\usepackage{hyperref}
\usepackage{amsthm}
\usepackage{amssymb}

\begin{document}
\def\be{\begin{equation}}
\def\ee{\end{equation}}

\def\bc{\begin{center}}
\def\ec{\end{center}}
\def\bea{\begin{eqnarray}}
\def\eea{\end{eqnarray}}
\newcommand{\avg}[1]{\langle{#1}\rangle}
\newcommand{\Avg}[1]{\left\langle{#1}\right\rangle}

\def\ie{\textit{i.e.}}
\def\etal{\textit{et al.}}
\def\m{\vec{m}}
\def\G{\mathcal{G}}

\newcommand{\davide}[1]{{\bf\color{blue}#1}}
\newcommand{\gin}[1]{{\bf\color{green}#1}}

\title{Short-time large deviation of constrained random acceleration process}
\author{Hanshuang Chen}\email{chenhshf@ahu.edu.cn}
\author{Lulu Tian}
\author{Guofeng Li}
\affiliation{School of Physics and Optoelectronic Engineering, Anhui University, Hefei 230601, China}
\begin{abstract}
We consider the motion of a randomly accelerated particle in one
dimension according to the Langevin equation $\ddot x(t)=\sqrt{2D}\xi(t)$, where $x(t)$ is the particle’s position, $\xi(t)$ is Gaussian white noise with zero mean, and $D$ is the particle velocity diffusion constant. By employing the optimal fluctuation method, we study the short-time distribution $P(\mathcal{A}=A) $ of the functionals, $\mathcal{A}=\int_{0}^{t_f} x^n(t) dt$, along constrained trajectories for a given time duration $t_f$, where $n$ is a positive integer. We consider two types of constraints: one is called the total constraint, where the initial position and velocity and the final position and velocity are both fixed, and the other is called the partial constraint, where the initial position and velocity, the final position are fixed, and letting the final velocity be free. Via the variation of constrained action functionals, the resulting Euler-Lagrange equations are analytically solved for $n=1$ and 2, and the optimal path, i.e., the most probable realization of the random acceleration process $x(t)$, conditioned on specified $A$ and $n$, are correspondingly obtained. For $n \geq 3$, a numerical scheme is proposed to find the optimal path. We show that, for $n=1$, $P(A)$ is a Gaussian distribution with the variance proportional to $Dt_f^5$. For $n \geq 2$, $P(A)$ exhibits the non-Gaussian feature. In the small-$A$ limit, $P(A)$ show a essential singularity,  $-\ln P(A) \sim A^{-3}$, and the optimal path localizes around the initial state over a long-time window, and then escapes to the final position sharply at a late time. For $A$ much larger than its typical value, there are multiple optimal paths with the same $A$ but with different actions (or probability densities). Among these degenerate paths, one with the minimum action is dominant, and the others are exponentially unlikely. All the theoretical results are validated by simulating the effective Langevin equations governing the constrained random acceleration process.

\end{abstract}

\maketitle
\section{Introduction}
The random acceleration process (RAP) is governed by the Langevin equation \cite{burkhardt2014first}
\begin{eqnarray}\label{eq1.0}
\ddot x(t)=\sqrt{2D}\xi(t),
\end{eqnarray} 
where $D$ is a constant. This equation describes the position of a particle in one dimension subject to a random force $\xi(t)$, where $\xi(t)$ is the Gaussian white noise with  $\langle {\xi (t)} \rangle  = 0$ and $\langle {\xi (t)}{\xi (t')} \rangle  = \delta ( {t - t'} )$. This is a non-Markov process as indicated by the second-order time derivative in Eq.(\ref{eq1.0}). However if one defines velocity $v=\dot x$, then the process becomes Markovian in the phase space $\left(x, v \right) $. Eq.(\ref{eq1.0}) can now be rewritten as,
\begin{eqnarray}\label{eq1.01}
\dot x(t)=v(t), \qquad  \dot v(t)=\sqrt{2D}\xi(t).
\end{eqnarray}

In recent years, the study of RAP has received growing attention, not only because the RAP serves as a representative of non-Markovian processes, but also because it has found wide applications in diverse areas of physics. Since the pioneering works of McKean \cite{mckean1962winding} and Marshall and Watson \cite{marshall1985drop},  first-passage properties of RAP and some related problems have been extensively studied, including persistence exponents \cite{bray2013persistence}, convex hull \cite{hilhorst2008sylvester,reymbaut2011convex}, extreme-value statistics, the distribution of time to reach maximum displacement \cite{majumdar2010time}, last passage time distribution \cite{lachal1994last}, residence time statistics \cite{boutcheng2016occupation,burkhardt2017occupation}, exit time statistics \cite{masoliver1996exact,kotsev2005randomly,burkhardt2007random},stochastic resetting \cite{singh2020random,capala2021random}, record statistics \cite{godreche2022record},
tagged particle statistics in single file process \cite{burkhardt2019tagged}, functional distributions \cite{sinai1992distribution}, partial
survival probability \cite{burkhardt2000dynamics,de2001partial}. The first-passage properties of RAP is closely related to semi-flexible polymer in a narrow cylindrical channel \cite{burkhardt1993semiflexible,burkhardt1997free}. Other applications include the spatial persistence of fluctuating interface with $z = 4$ \cite{PhysRevLett.86.3700}, crack propagation in elastic media \cite{schwarz2001first}, statistical properties of Burgers equation with Brownian initial velocity \cite{valageas2009statistical}.

In the free space, the propagator is expressed by multivariate Gaussian distribution \cite{burkhardt2014first}, 
\begin{eqnarray}\label{eq1.02}
P\left( {x,v,t|{x_0},{v_0}} \right) = \mathcal{N}\exp \left( { - \frac{1}{2}{\textbf{X}^{\top}}{C^{ - 1}}\textbf{X}} \right),
\end{eqnarray} 
where $\mathcal{N}=1/(2 \pi \sqrt{\det C} ) $ is the normalization constant, $\textbf{X}=\left(x-\langle {x} \rangle, v-\langle {v} \rangle \right)^{\top} $, and $C$ is the covariance matrix given by
\begin{eqnarray}\label{eq1.03}
C = \langle  \textbf{X} \textbf{X}^{\top} \rangle  = \left( {\begin{array}{*{20}{c}}
{2D{t^3}/3}&{D{t^2}} \\ 
{D{t^2}}&{2Dt} 
\end{array}} \right).
\end{eqnarray} 
Inserting the normalization constant and Eq.(\ref{eq1.03}) into Eq.(\ref{eq1.02}), one obtains
\begin{eqnarray}\label{eq1.04}
P( {x,v,t|{x_0},{v_0}} ) = \frac{{\sqrt 3 }}{{2\pi D{t^2}}} e^  { - \frac{3}{{D{t^3}}} \left[  {\left( {x - {x_0} - vt} \right)\left( {x - {x_0} - {v_0}t} \right) + \frac{{{t^2}}}{3}{{\left( {v - {v_0}} \right)}^2}} \right] } .
\end{eqnarray} 
Marginalization over $v$ one gets the distribution of the position $x$,
\begin{eqnarray}\label{eq1.05}
P( {x,t|{x_0},{v_0}} ) = \int_{ - \infty }^\infty  {dvP\left( {x,v,t|{x_0},{v_0}} \right)}  = \sqrt {\frac{3}{{4\pi D{t^3}}}} e^  { - \frac{3}{{4D{t^3}}}{{\left( {x - {v_0}t - {x_0}} \right)}^2}}  .
\end{eqnarray}

Using the property of the Gaussian noise in the Langevin equation (\ref{eq1.0}), it is clear that the probability of any path $\left\lbrace x(t)\right\rbrace $ with $0\leq t \leq T$ can be written as \cite{majumdar2007brownian}
\begin{eqnarray}\label{eq1.1}
P[\left\lbrace x(t)\right\rbrace] \propto  e^{-{S}},
\end{eqnarray} 
where 
\begin{eqnarray}\label{eq1.2}
{S}=\frac{1}{4D} \int_{0}^{t_f} \ddot x^2(t) dt
\end{eqnarray} 
is called the action along the stochastic path $\left\lbrace x(t)\right\rbrace $.
The propagator in Eq.(\ref{eq1.04}) can be thus written as the summation over the contributions from all possible paths propagating from $\left(x_0, v_0 \right)$ at $t=0$ to $\left(x, v\right)$ at $t=t_f$. This sum is the path integral representation of the propagator
\begin{eqnarray}\label{eq1.3}
P\left( {x,v,t_f|{x_0},{v_0}} \right) = \int_{\left( {{x_0},{v_0}} \right)}^{\left( {x,v} \right)} {\mathcal{D}x( t )} P[\left\lbrace x(t)\right\rbrace].
\end{eqnarray}

In the presence of constraints, the process is pushed into a large-deviation regime, and the action \ref{eq1.2} becomes very large in the short-time limit (or equivalent to the limit of $D\to 0$ in mathematics). Therefore, the dominating contribution to the probability distribution comes from the optimal path: a single deterministic
trajectory $x^{*}(t)$ which minimizes the action functional \ref{eq1.2} subject to the specified additional constraints. The minimization procedure leads to the Euler-Lagrange equation for the optimal path. Once the optimal path
$x^{*}(t)$ is determined, one can evaluate the probability distribution of the specific large deviation up to a pre-exponential factor,
\begin{eqnarray}\label{eq1.4}
-\ln P \simeq S[x^{*}(t)]
\end{eqnarray} 
by plugging the optimal path into the action functional $S[x(t)]$ given by Eq.(\ref{eq1.2}). This is the essence of the optimal fluctuation method (OFM), which traces back to early literature \cite{halperin1966impurity,zittartz1966theory,lifshitz1968theory}, and has found numerous application in different areas of physics. Recently, the OFM has been developed and applied in a number of studies
of large deviations of Brownian motion with various constraints \cite{meerson2019large,meerson2019geometrical,agranov2020airy,meerson2020area,majumdar2020statistics,nickelsen2018anomalous,smith2023anomalous,bar2023geometrical} and others \cite{meerson2016large,majumdar2020toward,meerson2022geometrical,meerson2023geometrical}.Interestingly, the OFM essentially becomes geometrical optics in the case of Brownian motion, where the optimal path is a geodesic subject to imposed constraints.

In a recent work, Meerson \cite{meerson2023geometrical} developed the OFM to study the large deviation properties of RAP. The author focused the $A \to 0$ tail of distribution of $P_n(A|L)$ of the functionals ${A}=\int_{0}^{T} x^n(t)dt$ along first-passage trajectories $x(t)$, starting from a position $L$ and terminating whenever hitting the origin at a random time $T$. The author found that
this tail has a universal essential singularity, $P_n (A \to 0|L)  \sim \exp \left( -\frac{\alpha_n L^{3n+2}}{DA^3} \right) $, where $\alpha_n$ is an $n$-dependent number was calculated analytically for $n=0$, 1, and 2 and numerically for other $n$.

In this work, we aim to study short-time large deviations of the functionals ${A}=\int_{0}^{t_f} x^n(t)dt$ along constrained trajectories of RAP with a fixed time duration $t_f$. In details, we consider two different ensembles of the constrained trajectories. The first one is called the total constraint, where the initial position and velocity and the final position and velocity are both fixed. The second one is called the partial constraint, where the initial position and velocity, the final position are fixed, and letting the final velocity be free. By the variation of the constrained action functionals, we obtain the resulting Euler-Lagrange equation. The optimal path $x^*(t)$, i.e., the most probable realization of the constrained RAP, can be analytically obtained for $n=1$ and $n=2$, by solving the Euler-Lagrange equation subject to boundary conditions. For $n \geq 3$, we provide a numerical scheme to find the optimal path. For $n=1$, the distribution $P(A)$ of $A$ exhibits a Gaussian function. For $n\geq 2$, $P(A)$ shows the non-Gaussian feature. In the limit of $A\to 0$, $-\ln P(A) \sim A^{-3}$, which shares a similar singularity with the previous result in Ref.\cite{meerson2023geometrical}. Interestingly, the optimal path for $A\to 0$ begins to localize around the initial state for a long time, and then leaps sharply to the final state at a late time. To numerical validate our theoretical results, we have derived the effective Langevin equations for generating the two types of the constrained trajectories. All results show the excellent agreement between the theory and simulations.

\section{Optimal paths for the constrained RAP}
In the present work, we consider two different ensembles of constrained trajectories of RAP with a fixed time duration $t_f$. For both cases, the trajectories starts from a fixed initial position $x(0)=x_0$ and a fixed velocity $v(0)=v_0$. The constraints in the final state are different. For the first case (\textit{total} constraint), the final position $x(t_f)=x_f$ and velocity $v(t_f)=v_f$ are both fixed. For the second case (\textit{partial} constraint), the final position is fixed, while the final velocity is not fixed. As mentioned before, in the short-$t_f$ limit or in the limit of $D \to 0$, the probability distribution of constrained trajectories is dominated by the optimal path with a minimum action. Via the minimization of action in Eq.(\ref{eq1.2}), it leads to the Euler-Lagrange equation
(see appendix \ref{appendix1} for the details of derivation)
\begin{eqnarray}\label{eq1.6.0}
x^{(4)}(t)=0,
\end{eqnarray}  
where the superscript (4) denotes the fourth derivative with
respect to time.

\subsection{Total constraint: $x_f$ and $v_f$ are both fixed}
For total constraint, the starting position and velocity, and the ending position and velocity are fixed, i.e., 
\begin{eqnarray}\label{eq1.1.0}
x(0)=x_0, \quad v(0)=v_0, \quad x(t_f)=x_f, \quad v(t_f)=v_f.
\end{eqnarray} 
The optimal path $x^{*}(t)$ can be obtained by solving the Euler-Lagrange equation \ref{eq1.6.0} subject to the boundary conditions in Eq.(\ref{eq1.1.0}), which yields
\begin{eqnarray}\label{eq1.7.0}
x^{*}(t) =  {x_0} + {v_0}t - \frac{{2{v_0}{t_f} + {v_f}{t_f} + 3{x_0} - 3{x_f}}}{{t_f^2}}{t^2}  + \frac{{{v_0}{t_f} + {v_f}{t_f} + 2{x_0} - 2{x_f}}}{{t_f^3}}{t^3} .
\end{eqnarray} 
The corresponding minimum action is 
\begin{eqnarray}\label{eq1.8.0}
{S}_{\min}= \frac{ {v_0^2 + {v_0}{v_f} + v_f^2}}{D t_f} +\frac{3 ( {{v_0} + {v_f}} )\left( {{x_0} - {x_f}} \right) + 3{{( {{x_0} - {x_f}} )}^2}}{D t_f^2}.
\end{eqnarray}

\subsection{Partial constraint: $x_f$ is fixed and $v_f$ is free}
For the partial constraint, we relax one of the constraints such that the final velocity $v_f$ is a free parameter. Such a setting can bring an extra boundary condition (see Eq.(\ref{eqa1.5}) in the appendix \ref{appendix1} for details). All the boundary conditions are summarized as follows.   
\begin{eqnarray}\label{eq1.7}
x(0)=x_0, \quad v(0)=v_0, \quad x(t_f)=x_f, \quad \ddot x(t_f)=0.
\end{eqnarray}  
By solving the Euler-Lagrange equation under the  boundary conditions in Eq.(\ref{eq1.7}), we obtain the optimal path
\begin{eqnarray}\label{eq1.10}
x^{*}(t)=x_0+v_0 t+\frac{{3\left( {x_f - {v_0}t_f - {x_0}} \right)}}{{2{t_f^2}}} t^2 - \frac{{x_f - {v_0}t_f - {x_0}}}{{2{t_f^3}}} t^3,
\end{eqnarray}  
and the corresponding action  
\begin{eqnarray}\label{eq1.11}
{S}_{\min}=\frac{3(x_f-v_0 t_f-x_0)^2}{4D t_f^3}.
\end{eqnarray}  
The final velocity at $t=t_f$ along the optimal path can be obtained by taking the derivative of Eq.(\ref{eq1.10}) with respect to $t$,
\begin{eqnarray}\label{eq1.12}
v^{*}(t_f)=v_0+\frac{3(x_f-v_0 t_f-x_0)}{2t_f}.
\end{eqnarray}  
Eq.(\ref{eq1.12}) can be also verified by extremizing Eq.(\ref{eq1.04}) with respect to $v$. We then insert the most likely final velocity $v{*}(t_f)$ into the exponential factor of $P( {x=x_f,v,t=t_f|{x_0},{v_0}} )$ in Eq.(\ref{eq1.04}) and thus obtain the minimum action in Eq.(\ref{eq1.11}).

\section{Effective Langevin equations for constrained RAP}
It is inefficient to generate the stochastic processes conditioned on the final state from directedly simulating the Langevin equation \ref{eq1.01}. Fortunately, Majumdar and Orland proposed a novel method to 
generate the constrained paths by deriving the effective Langevin equation \cite{majumdar2015effective}. For completeness, we outline the main result in the appendix \ref{appendix2}, and we simultaneously generalize the result to more general situations where partial constraints is present in high-dimensional systems.

\begin{figure}
	\centerline{\includegraphics*[width=0.8\columnwidth]{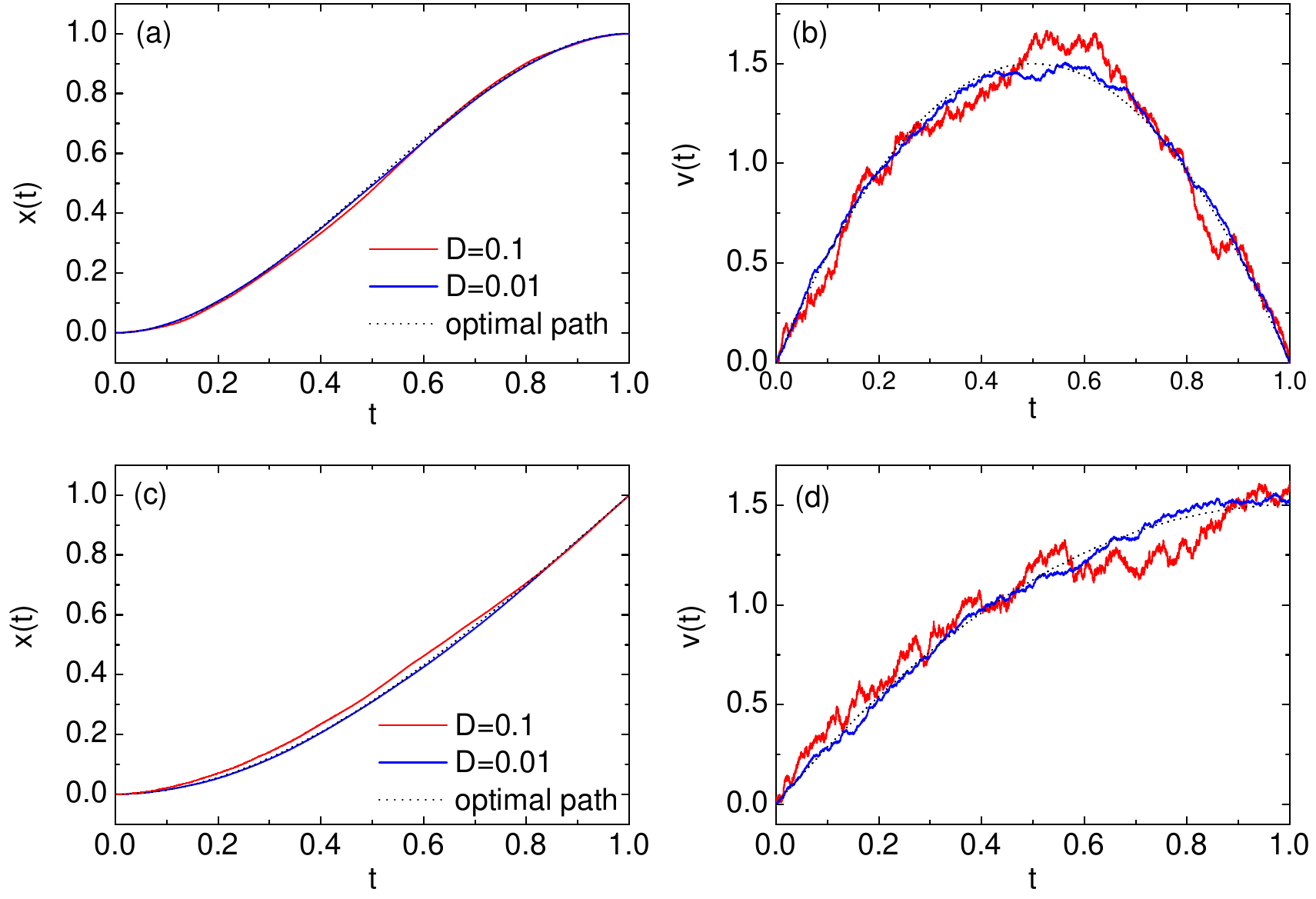}}
	\caption{Totally (a-b) and partially (c-d) constrained trajectories of random acceleration process  generated by the effective Langevin equations (\ref{eq.a7}) and (\ref{eq.a9}) for two different $D$: $D=0.01$ and $D=0.1$. For the total constraint, $x_0=v_0=0$, $x_f=1$, $v_f=0$, and $t_f=1$. For the partial constraint, $x_0=v_0=0$, $x_f=1$, $t_f=1$, and $v_f$ is a free parameter. A comparison is done by plotting the optimal path Eq.(\ref{eq1.7.0}) and Eq.(\ref{eq1.10}) with the minimum action. (a,c) and (b,d) correspond to the time evolutions of $x(t)$ and $v(t)$, respectively. \label{fig1}}
\end{figure}

Following the result in the appendix \ref{appendix2}, for the RAP we have $\mathbf{X}=(x, v )^{\top}$, $\mathbf{u}=(v, 0)^{\top}$, $\sigma=(0, \sqrt{2D} )^{\top} $, and thus $B=\sigma \sigma^{\top}=\left( \begin{array}{*{20}{c}}
{0}&{0} \\ 
{0}&{2D} 
\end{array}\right) $. Firstly, we consider a stochastic path of a fixed time duration $t_f$ where the initial state and final state are both fixed, i.e., $\left(x(0) ,v(0) \right)=\left(x_0, v_0 \right)  $ and $\left(x(t_f) ,v(t_f) \right)=\left(x_f, v_f \right)  $. The backward propagator $Q(x,v,t)=P(x_f,v_f,t_f|x,v,t)$ can be obtained from Eq.(\ref{eq1.04}), and it is not hard to write the effective Langevin equation for the totally constrained RAP,
\begin{eqnarray}\label{eq.a7}
\left\{\begin{gathered}
\dot{x_b} = {v_b} \hfill \\
\dot{v_b} =  \frac{{6( {{x_f} - x_b} )}}{{{{( {t - {t_f}} )}^2}}} + \frac{{6{v_f} + 4( {v_b - {v_f}} )}}{{t - {t_f}}}  + \sqrt {2D} \xi(t) \hfill \\ 
\end{gathered}  \right.
\end{eqnarray}
In the limit of $D \to 0$, Eq.(\ref{eq.a7}) reduces to the deterministic differential equations, whose solution yields the optimal path $x^{*}(t)$ in Eq.(\ref{eq1.7.0}). In Fig.\ref{fig1}(a-b), we show two typical trajectories for the totally constrained RAP  generated by the effective Langevin equation (\ref{eq.a7}) for two different values of $D$: $D=0.1$ and $D=0.01$. For comparison, we also plot the optimal path with the minimum action as shown in Eq.(\ref{eq1.7.0}). Obviously, the optimal path is more matching with the stochastic trajectory with a smaller value of $D$.

Secondly, we consider the case of partial constraint such that the final velocity $v_f$ is relaxed to be a free parameter. Under this case, $Q(x,v,t)=\int d{v_f} P(x_f,v_f,t_f|x,v,t)$ can be obtained from Eq.(\ref{eq1.05}), and the effective Langevin equation for the partially constrained RAP reads
\begin{eqnarray}\label{eq.a9}
\left\{\begin{gathered}
\dot{x_b} = {v_b} \hfill \\
\dot{v_b} =\frac{{3\left[ {{x_f} - x_b - v_b\left( {{t_f} - t} \right)} \right]}}{{{{( {{t_f} - t} )}^2}}} + \sqrt {2D} \xi(t) \hfill \\ 
\end{gathered} \right.
\end{eqnarray}

As expected, the optimal path $x^{*}(t)$ in Eq.(\ref{eq1.10}) corresponds to the  solution of Eq.(\ref{eq.a9}) in the case of $D=0$.  In Fig.\ref{fig1}(c-d), we show two  trajectories of the partially constrained RAP generated by the effective Langevin equation (\ref{eq.a9}) with $D=0.1$ and $D=0.01$. We also plot the optimal path shown in Eq.(\ref{eq1.10}). It is clear that the optimal path is a more realistic characterization of the noisy trajectory with a smaller value of $D$.

\section{Distribution of functionals along the constrained RAP}
In this section, we study the short-time large deviations of the functionals along totally (or partially) constrained stochastic trajectories of RAP. The functionals are defined as
\begin{eqnarray}\label{eq3.1}
\mathcal{A}=\int_{0}^{t_f} x^n(t) dt,
\end{eqnarray}
where $n$ is a positive integer. For $n=1$, the functionals gives the area swept by $x(t)$ in a fixed time $t_f$. The case $n=2$ corresponds to the statistics of the moment
of inertia of a semiflexible polymer chain of a given length in narrow channels.
The main goal in this section is to find the distribution $P(\mathcal{A}=A)$ of the functionals for arbitrary $n$ by using the OFM. The method is justified in the short-time limit $t_f \to 0$ (or in the low-noise limit $D \to 0$). 

The minimization of the action subject to the constraint (\ref{eq3.1}) can be achieved by minimizing the modified action functionals
\begin{eqnarray}\label{eq3.2}
\mathcal{S}_{\lambda}=\frac{1}{2D}\int_{0}^{t_f} \left[  \frac{1}{2} \ddot x^2(t) -\lambda x^n(t) \right]  dt,
\end{eqnarray}
where $\lambda$ is the Lagrange multiplier that enforces the constraint $\mathcal{A}=A$. Via the variation of the constrained action functionals, the resulting Euler-Lagrange equation reads (see Eq.(\ref{eqa1.5}))
\begin{eqnarray}\label{eq3.3}
x^{(4)}(t)-\lambda n x^{n-1}(t)=0.
\end{eqnarray} 
 
\subsection{$n=1$}
For $n = 1$, the Euler-Lagrange equation (\ref{eq3.3}) is very simple: $x^{(4)}(t)=\lambda $. Its solution is a quartic parabola.  
\begin{eqnarray}\label{eq3.3.1}
x^{*}( t ) = {x_0} + {v_0}t + c_2 {t^2}  + c_3 {t^3} + \frac{\lambda }{{24}}{t^4},
\end{eqnarray} 
where we have used the boundary conditions $x(0)=x_0$ and $v(0)=v_0$. Using the other two boundary conditions in Eq.(\ref{eq1.1.0}) and Eq.(\ref{eq1.7}) for the total constraint and partial constraint, respectively, we determine the two unknown coefficients in Eq.(\ref{eq3.3.1}), 
\begin{eqnarray}\label{eq3.3.2}
c_2 =\left\{ \begin{array}{lll} \frac{{\lambda t_f^4 + 72\left( {{x_f} - {x_0}} \right) - 24{t_f}\left( {2{v_0} + {v_f}} \right)}}{{24t_f^2}},    &{\rm{for}} \, {\rm{total}} \, {\rm{constraint}} ,  \\
\frac{{\lambda t_f^4 + 24\left( {{x_f} - {x_0}} \right) - 24{v_0}{t_f}}}{{16t_f^2}},    & {\rm{for}} \, {\rm{partial}} \, {\rm{constraint}},  
\end{array}  \right. 
\end{eqnarray}
and
\begin{eqnarray}\label{eq3.3.3}
c_3 =\left\{ \begin{array}{lll} \frac{{   24\left( {{x_0} - {x_f}} \right) + 12{t_f}\left( {{v_0} + {v_f}} \right)} - \lambda t_f^4}{{12t_f^3}},    & {\rm{for}} \, {\rm{total}} \, {\rm{constraint}} ,  \\
- \frac{{5\lambda t_f^4 + 24\left( {{x_f} - {x_0}} \right) - 24{v_0}{t_f}}}{{48t_f^3}},    & {\rm{for}} \, {\rm{partial}} \, {\rm{constraint}}.
\end{array}  \right. 
\end{eqnarray}

The Lagrange multiplier $\lambda$ in Eq.(\ref{eq3.3.1}) is determined by the constraint $A=\int_{0}^{t_f} x(t) dt$, yielding
\begin{eqnarray}\label{eq3.3.4}
\lambda =\left\{ \begin{array}{lll} \frac{{60\left[ {12A + t_f^2\left( {{v_f} - {v_0}} \right) - 6{t_f}\left( {{x_0} + {x_f}} \right)} \right]}}{{t_f^5}},    & {\rm{for}} \, {\rm{total}} \, {\rm{constraint}} ,  \\
\frac{{40\left( {8A - {v_0}t_f^2 - 5{x_0}{t_f} - 3{x_f}{t_f}} \right)}}{{t_f^5}},    & {\rm{for}} \, {\rm{partial}} \, {\rm{constraint}},  .
\end{array}  \right. 
\end{eqnarray}
Substituting Eq.(\ref{eq3.3.1}) and Eq.(\ref{eq3.3.4}) into Eq.(\ref{eq1.2}), we obtain the action along the optimal path and the distribution of $\mathcal{A}=A$, 
\begin{eqnarray}\label{eq3.3.5}
-\ln P(A)  \simeq    \frac{( {A - {A_p}} )^2}{2 \sigma_A^2},
\end{eqnarray} 
where
\begin{eqnarray}\label{eq3.3.6}
A_p =\left\{ \begin{array}{lll} \frac{{{t_f}\left[ {{t_f}\left( {{v_0} - {v_f}} \right) + 6\left( {{x_0} + {x_f}} \right)} \right]}}{{12}},    & {\rm{for}} \, {\rm{total}} \, {\rm{constraint}} ,  \\
\frac{{{t_f}\left( {{v_0}{t_f} + 5{x_0} + 3{x_f}} \right)}}{8},    & {\rm{for}} \, {\rm{partial}} \, {\rm{constraint}},  
\end{array}  \right. 
\end{eqnarray}
is the most likely value (or mean value) of the functionals $\mathcal{A}$, and
\begin{eqnarray}\label{eq3.3.7}
\sigma_A^2 =\left\{ \begin{array}{lll} Dt_f^5/360,    & {\rm{for}} \, {\rm{total}} \, {\rm{constraint}} ,  \\
Dt_f^5/160,    & {\rm{for}} \, {\rm{partial}} \, {\rm{constraint}},
\end{array}  \right. 
\end{eqnarray}
is the variance of the functionals $\mathcal{A}$. The result in Eq.(\ref{eq3.3.5}) shows that the distribution of $\mathcal{A}=A$ is a Gaussian function for both cases. This is easily understood because that $x(t)$ satisfies a Gaussian distribution and the integral $\mathcal{A}=\int_{0}^{t_f} x(t) dt$ is linear which is also Gaussian. However, different constraints on the stochastic trajectories can lead to different expected values and variances. With less constraints, the variance will become larger, as expected.

\begin{figure}
	\centerline{\includegraphics*[width=0.8\columnwidth]{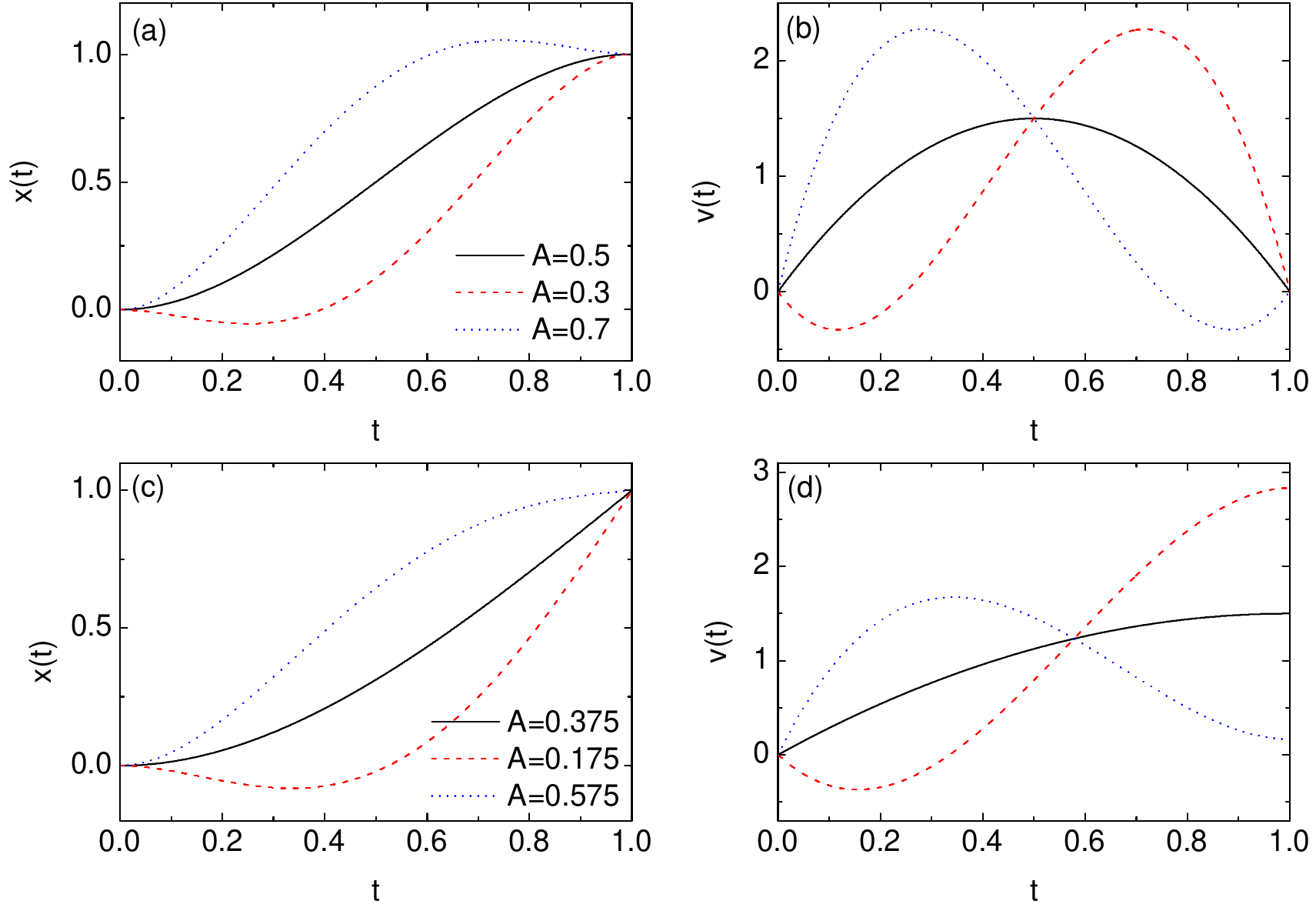}}
	\caption{The optimal paths for totally (a-b) and partially (c-d) constrained random acceleration process with a given value of the functionals $A=\int_{0}^{t_f} x(t) dt$. (a) and (c) correspond to the time evolution of displacement $x(t)$, (b) and (d) to the evolution of  velocity $v(t)$. For the case of total constraint (a-b), $x_0=v_0=0$, $x_f=1$, $v_f=0$, and $t_f=1$. For the case of partial constraint (c-d), $x_0=v_0=0$, $x_f=1$, $t_f=1$, and $v_f$ is a free parameter.    \label{fig4}}
\end{figure}

\begin{figure}
	\centerline{\includegraphics*[width=0.6\columnwidth]{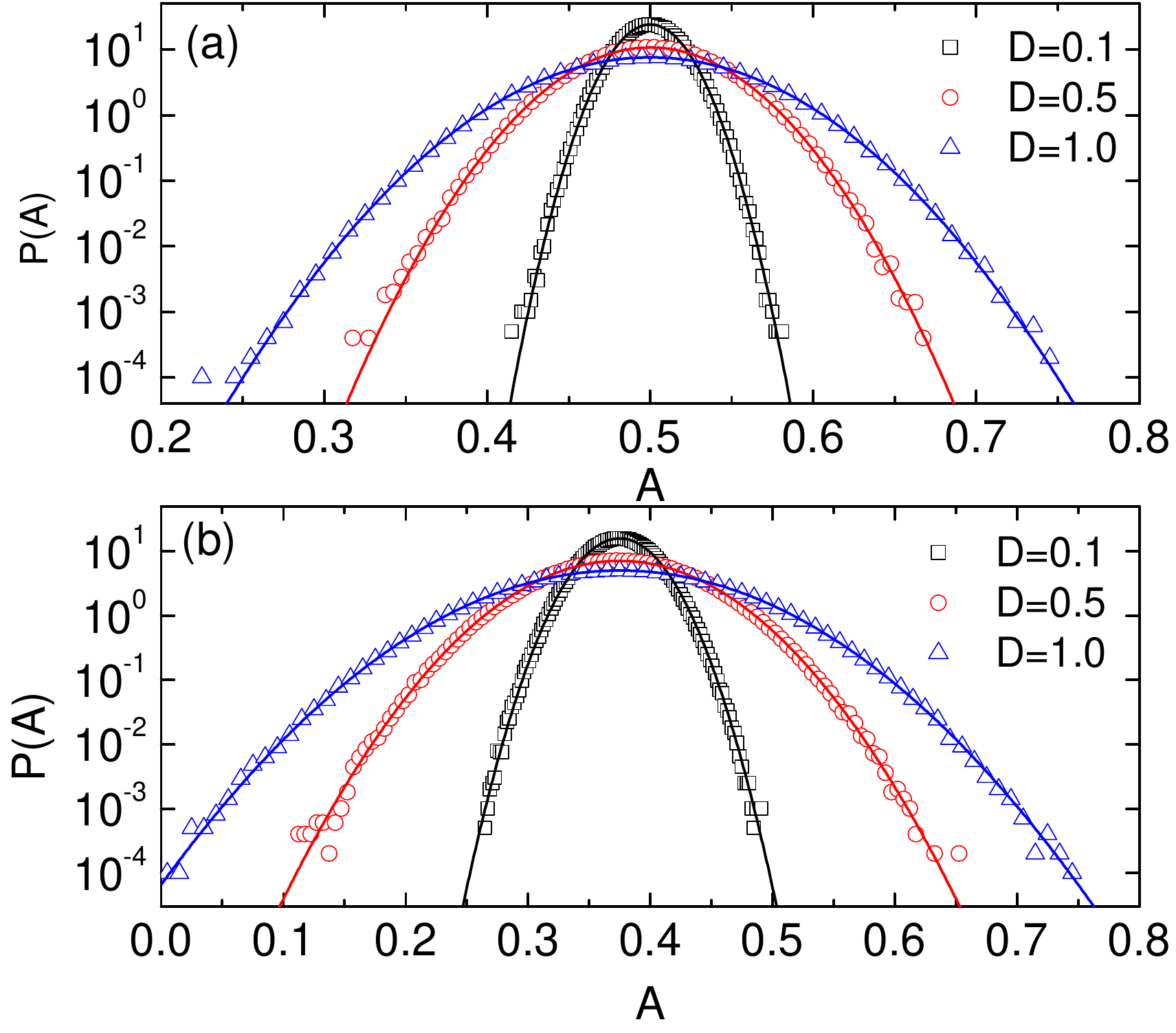}}
	\caption{Distribution of $A=\int_{0}^{t_f} x(t) dt$ for totally (a) and partially (b) constrained trajectories of random acceleration process for three different $D=0.1,0.5,1.0$. In (a), $x_0=v_0=0$, $x_f=1$, $v_f=0$, and $t_f=1$. In (b), $x_0=v_0=0$, $x_f=1$, $t_f=1$, and $v_f$ is a free parameter. Lines and symbols correspond to the theoretical and simulation results, respectively. \label{fig3}}
\end{figure}

In Fig.\ref{fig4}(a-b), we show the optimal path $x^{*}(t)$ and velocity $v^{*}(t)$ for totally constrained case with three different $A$, where $x_0=v_0=v_f=0$, $x_f=1$, and $t_f=1$. Under such a setting, $A_p=0.5$ is the most likely value of the functionals $\mathcal{A}$ in terms of Eq.(\ref{eq3.3.6}). In Fig.\ref{fig3}(a), we show the distribution of $\mathcal{A}$ for totally constrained paths with three different values of $D$. We have also performed simulations to verify the theoretical predictions. In simulations, we have collected $10^6$ independent trajectories generated by the effective Langevin equations \ref{eq.a7}. It is clear that the simulation results are consistent with the theory. 

In Fig.\ref{fig4}(c-d), we show the optimal path $x^{*}(t)$ and velocity $v^{*}(t)$ for partially constrained case with three different $A$, where $x_0=v_0=0$, $x_f=1$, and $t_f=1$. In this case, $A_p=0.375$ is the most likely value of the functionals $\mathcal{A}$ in terms of Eq.(\ref{eq3.3.6}). In Fig.\ref{fig3}(b), we show the distribution of $\mathcal{A}$ for partially constrained paths with three different values of $D$. Again, the agreement is achieved between theory and simulations.

\subsection{$n=2$}
For $n = 2$, the Euler-Lagrange equation (\ref{eq3.3}) becomes: $x^{(4)}(t)=2 \lambda x(t)  $, and the general solution of the equation can given by $\lambda>0$ and $\lambda<0$, separately,  
\begin{eqnarray}\label{eq4.1}
x\left( t \right) =\left\{ \begin{array}{lll}  {c_1}\cos \left( {\Lambda t} \right) + {c_2}\sin \left( {\Lambda t} \right) + {c_3}{e^{ - \Lambda t}} + {c_4}{e^{\Lambda t}},    & \lambda>0,  \\
\cos \left( {\Lambda t} \right)\left( {{c_1}{e^{\Lambda t}} + {c_2}{e^{ - \Lambda t}}} \right) + \sin \left( {\Lambda t} \right)\left( {{c_3}{e^{\Lambda t}} + {c_4}{e^{ - \Lambda t}}} \right),    & \lambda<0,  \\ 
\end{array}  \right. 
\end{eqnarray}
where $\Lambda>0$ is given by
\begin{eqnarray}\label{eq4.1.1}
\Lambda^4 =\left\{ \begin{array}{lll} 2 \lambda,    & \lambda>0,  \\
-\lambda/2,    & \lambda<0,  \\ 
\end{array}  \right. 
\end{eqnarray}
and four unknown coefficients $c_1$, $c_2$, $c_3$ and $c_4$ in Eq.(\ref{eq4.1}) can be  determined by the boundary conditions.

We first consider the case of total constraint: $x(0)=x_0$, $v(0)=v_0$, $x(t_f)=x_f$, $v(t_f)=v_f$. For simplicity, we let $x_0=v_0=v_f=0$. For $\lambda>0$, four coefficients are given by
\begin{eqnarray}\label{eq4.2}
\left\{ \begin{gathered}
{c_1} = \frac{{{x_f}\left[ {1 + {e^{2\Lambda {t_f}}} - 2{e^{\Lambda {t_f}}}\cos \left( {\Lambda {t_f}} \right)} \right]}}{{2\left( {1 + {e^{2\Lambda {t_f}}}} \right)\cos \left( {\Lambda {t_f}} \right) - 4{e^{\Lambda {t_f}}}}} \hfill \\
{c_2} = \frac{{{e^{\Lambda {t_f}}}{x_f}\left[ {\cos \left( {\Lambda {t_f}} \right) - {e^{\Lambda {t_f}}} - \sin \left( {\Lambda {t_f}} \right)} \right]}}{{2\left( {1 + {e^{2\Lambda {t_f}}}} \right)\cos \left( {\Lambda {t_f}} \right) - 4{e^{\Lambda {t_f}}}}} \hfill \\
{c_3} = \frac{{{x_f}\left[ {1 - {e^{2\Lambda {t_f}}} - 2{e^{\Lambda {t_f}}}\sin \left( {\Lambda {t_f}} \right)} \right]}}{{2\left( {1 + {e^{2\Lambda {t_f}}}} \right)\cos \left( {\Lambda {t_f}} \right) - 4{e^{\Lambda {t_f}}}}} \hfill \\
{c_4} = \frac{{{x_f}\left[ {{e^{\Lambda {t_f}}}\cos \left( {\Lambda {t_f}} \right) + {e^{\Lambda {t_f}}}\sin \left( {\Lambda {t_f}} \right) - 1} \right]}}{{2\left( {1 + {e^{2\Lambda {t_f}}}} \right)\cos \left( {\Lambda {t_f}} \right) - 4{e^{\Lambda {t_f}}}}} \hfill \\ 
\end{gathered}  \right.
\end{eqnarray} 
While for $\lambda<0$, four coefficients are given by
\begin{eqnarray}\label{eq4.2.1}
\left\{ \begin{gathered}
{c_1} = \frac{{{x_f}\left[ {\cos \left( {\Lambda {t_f}} \right)\left( {{e^{2\Lambda {t_f}}} - 1} \right) + \sin \left( {\Lambda {t_f}} \right)\left( {{e^{2\Lambda {t_f}}} + 1} \right)} \right]}}{{{e^{ - \Lambda {t_f}}} - 4{e^{\Lambda {t_f}}} + {e^{3\Lambda {t_f}}} + 2{e^{\Lambda {t_f}}}\cos \left( {2\Lambda {t_f}} \right)}} \hfill \\
{c_2} =  - \frac{{{x_f}\left[ {\cos \left( {\Lambda {t_f}} \right)\left( {{e^{2\Lambda {t_f}}} - 1} \right) + \sin \left( {\Lambda {t_f}} \right)\left( {{e^{2\Lambda {t_f}}} + 1} \right)} \right]}}{{{e^{ - \Lambda {t_f}}} - 4{e^{\Lambda {t_f}}} + {e^{3\Lambda {t_f}}} + 2{e^{\Lambda {t_f}}}\cos \left( {2\Lambda {t_f}} \right)}} \hfill \\
{c_3} =  - \frac{{{x_f}\sin \left( {\Lambda {t_f}} \right)\left[ {3 - {e^{2\Lambda {t_f}}} + \cos \left( {\Lambda {t_f}} \right)\left( {{e^{2\Lambda {t_f}}} - 1} \right)} \right]}}{{{e^{ - \Lambda {t_f}}} - 4{e^{\Lambda {t_f}}} + {e^{3\Lambda {t_f}}} + 2{e^{\Lambda {t_f}}}\cos \left( {2\Lambda {t_f}} \right)}} \hfill \\
{c_4} =  - \frac{{{x_f}\sin \left( {\Lambda {t_f}} \right)\left[ { - 1 + 3{e^{2\Lambda {t_f}}} + \cos \left( {\Lambda {t_f}} \right)\left( {{e^{2\Lambda {t_f}}} - 1} \right)} \right]}}{{{e^{ - \Lambda {t_f}}} - 4{e^{\Lambda {t_f}}} + {e^{3\Lambda {t_f}}} + 2{e^{\Lambda {t_f}}}\cos \left( {2\Lambda {t_f}} \right)}} \hfill \\ 
\end{gathered}  \right.
\end{eqnarray} 

Substituting the four coefficients in Eq.(\ref{eq4.2}) into Eq.(\ref{eq4.1}) and then performing the integral in Eq.(\ref{eq3.1}), we establish the relation between $A$ and $\Lambda$, 
\begin{eqnarray}\label{eq4.3}
A=f_1(\Lambda)
\end{eqnarray} 
with 
\begin{eqnarray}\label{eq4.4}
f_1(\tilde \Lambda) =\left\{ \begin{array}{lll} \frac{{x_f^2{t_f}}}{{8\tilde \Lambda {{(\cos \tilde \Lambda \cosh \tilde \Lambda  - 1)}^2}}}\left\{ \tilde \Lambda \left[ {2 + \cos \left( {2\tilde \Lambda } \right) + \cosh \left( {2\tilde \Lambda } \right)} \right] - 2\cosh \tilde \Lambda (2\tilde \Lambda \cos \tilde \Lambda  + 3\sin \tilde \Lambda ) \right.  \\ \left. + 3\sin \left( {2\tilde \Lambda } \right){{\cosh }^2}\tilde \Lambda   - 6\cos \tilde \Lambda \sinh \tilde \Lambda  + 3\sinh \left( {2\tilde \Lambda } \right){{\cos }^2}\tilde \Lambda  \right\},    & \lambda>0,  \\
\frac{{x_f^2{t_f}}}{{8\tilde \Lambda {{(\cos \left( {2\tilde \Lambda } \right) + \cosh \left( {2\tilde \Lambda } \right) - 2)}^2}}}\left\{ {12\left[ {\sin \left( {2\tilde \Lambda } \right) + \sinh \left( {2\tilde \Lambda } \right)} \right]\left[ {\cosh \left( {2\tilde \Lambda } \right) - 1} \right] - 32\tilde \Lambda {{\sin }^2}\tilde \Lambda {{\sinh }^2}\tilde \Lambda } \right\},    & \lambda<0,  \\ 
\end{array}  \right. 
\end{eqnarray} 
where $\tilde \Lambda= \Lambda t_f$.  However, Eq.(\ref{eq4.4}) is too complicate so that we cannot solve $\Lambda=f_1^{-1}(A)$ explicitly. 
The action defined in Eq.(\ref{eq1.2}) can be computed as
\begin{eqnarray}\label{eq4.4.2}
S(\tilde \Lambda) =\left\{ \begin{array}{lll} \frac{{{{\tilde \Lambda }^3}x_f^2}}{{32Dt_f^3{e^{2\tilde \Lambda }}{{\left( {\cos \tilde \Lambda \cosh \tilde \Lambda  - 1} \right)}^2}}}\left\{ \tilde \Lambda \left[ {2 + \cos \left( {2\tilde \Lambda } \right) + \cosh \left( {2\tilde \Lambda } \right)} \right] + 2\cosh \tilde \Lambda \left[ {\sin \tilde \Lambda  - 2\tilde \Lambda \cos \tilde \Lambda } \right] \right.  \\ \left. - \sin \left( {2\tilde \Lambda } \right){{\cosh }^2}\tilde \Lambda  + 2\cos \tilde \Lambda \sinh \tilde \Lambda  - \sinh \left( {2\tilde \Lambda } \right){{\cos }^2}\tilde \Lambda  \right\},    & \lambda>0,  \\
\frac{{{{\tilde \Lambda }^3}x_f^2}}{{4Dt_f^3{{\left[ {\cos \left( {2\tilde \Lambda } \right) + \cosh \left( {2\tilde \Lambda } \right) - 2} \right]}^2}}}\left\{ {2\left[ {\sin \left( {2\tilde \Lambda } \right) + \sinh \left( {2\tilde \Lambda } \right)} \right]\left[ {\cosh \left( {2\tilde \Lambda } \right) - 1} \right] + 16\tilde \Lambda {{\sin }^2}\tilde \Lambda {{\sinh }^2}\tilde \Lambda } \right\},    & \lambda<0,   
\end{array}  \right. 
\end{eqnarray}

For $\lambda=0$, one obtains that the typical value of $A=A_p$ with 
\begin{eqnarray}\label{eq4.4.1}
A_p= \frac{{13x_f^2{t_f}}}{{35}}.
\end{eqnarray} 
The corresponding path defined in Eq.(\ref{eq1.7.0}) possesses the minimum action.

\begin{figure}
	\centerline{\includegraphics*[width=0.8\columnwidth]{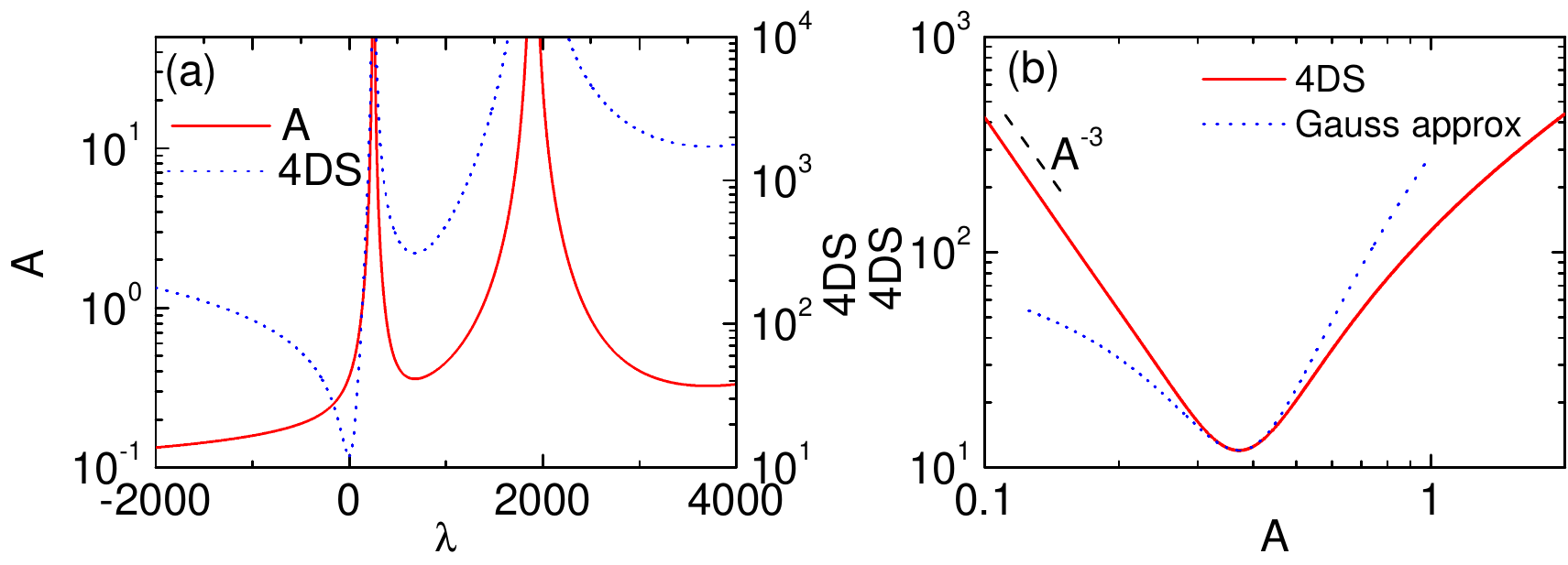}}
	\caption{Total constraint with $n=2$. (a) The functionals $A=\int_{0}^{t_f}x^2(t)dt$ (solid line) and the action $4DS$ (dotted line) as a function of the Lagrange multiplier $\lambda$. (b) The parametric plot of $4DS$ as a function of $A$ (solid line). The dotted line shows the Gaussian approximation. The other parameters are: $x_0=v_0=0$, $x_f=1$, $v_f=0$, and $t_f=1$. \label{fig5}}
\end{figure}

In Fig.\ref{fig5}(a), we plot $A$ and $S$ as a function of $\lambda$ for $x_f=t_f=1$. We see that, for $\lambda<0$, $A$ changes monotonically with $\lambda$, and $A$ tends to zero as $\lambda \to -\infty$. Correspondingly, the optimal path with a smaller $\lambda$ has a larger $S$, i.e., with a smaller probability. The asymptotic behavior of $P(A)$ in the small-$A$ limit can be order. In the limit of $\lambda \to -\infty$, $A \sim 2^{4/3}x_f^2 / |\lambda|^{1/4}$ and $S \sim x_f^2 |\lambda|^{3/4} /( 2^{11/4}D )$, which leads to a power-law divergence of $S$ as $A \to 0$ (as shown in Fig.\ref{fig5}(b)), 
\begin{eqnarray}\label{eq4.8}
-\ln P(A) \sim   A^{-3}, \quad A \to 0. 
\end{eqnarray}

On the other hand, for $\lambda>0$, $A$ and $S$ show more complex variations with  $\lambda$. At some values of $\lambda=\lambda_1, \lambda_2, \cdots$, $A$ and $S$ diverge, which $\lambda_1<\lambda_2<\cdots$ are determined by the roots of the equation $\cos \left( {{(2 \lambda)^{1/4} }{t_f}} \right)\cosh \left( {{(2 \lambda)^{1/4}}{t_f}} \right) = 1$. For a given $A$, there may be multiple $\lambda$ corresponding to different optimal paths. The actions along these degenerate paths are significantly different. One can identify the optimal path with the smallest action as the dominant one, while the others are exponentially unlikely.  In general, the dominant path is determined by the smallest value of $\lambda$. This is because that $S$ shows an up-stair shape with $\lambda$, and $S$ changes more sharply as $\lambda \to \lambda_i$ from left than right. In Fig.\ref{fig5}(b), we show a parametric plot of $S$ as a function of $A$, where the parameter $\lambda$ is bounded on $\lambda<\lambda_1\approx 250.282 $ for $x_f=t_f=1$.

\begin{figure}[h]
	\centerline{\includegraphics*[width=0.8\columnwidth]{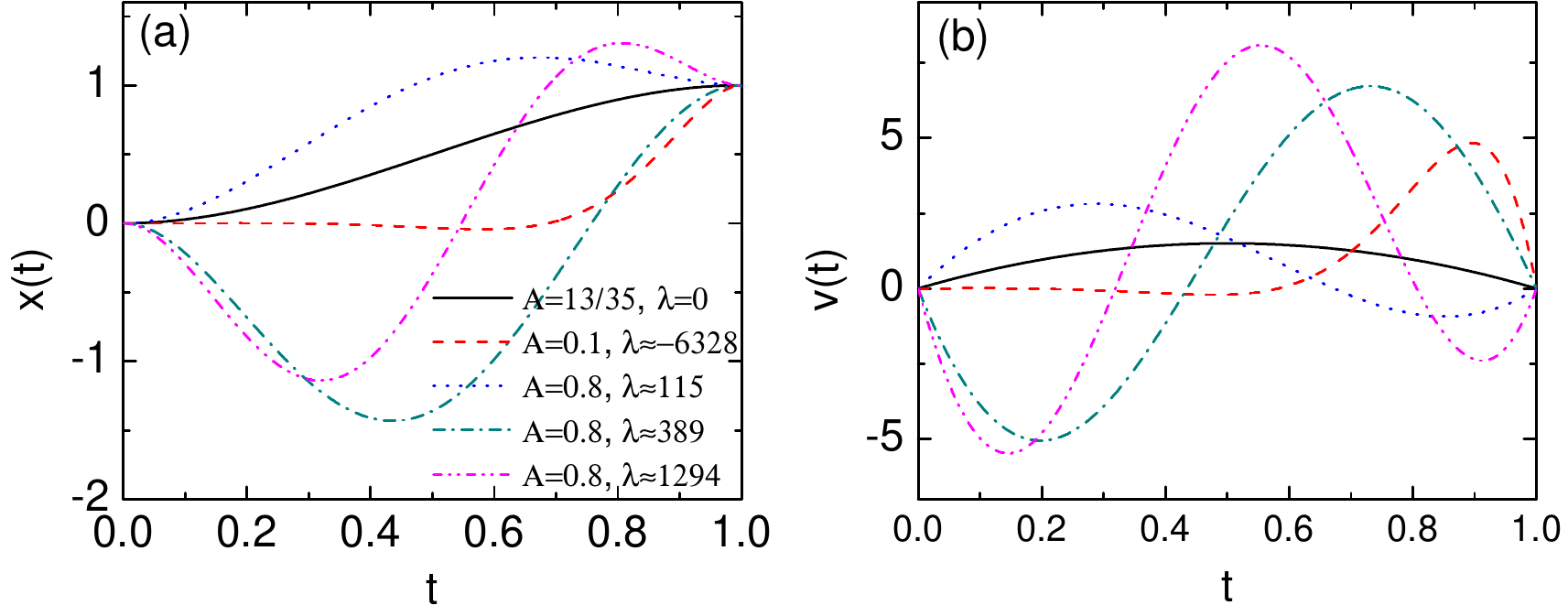}}
	\caption{Total constraint with $n=2$. The optimal path $x(t)$ (a) and the optimal velocity $v(t)$ for totally constrained trajectories of random acceleration process for different values of the functionals $A=\int_{0}^{t_f} x^2(t) dt$. The other parameters are: $x_0=v_0=0$, $x_f=1$, $v_f=0$, and $t_f=1$.  \label{fig6}}
\end{figure}

In Fig.\ref{fig6}, we show the optimal path for three different $A$: $A=A_p=13/35$, $A=0.1$ and $A=0.8$, where $x_f=t_f=1$. For $A=A_p$, $x^{*}(t)$ is a cubic curve as shown in Eq.(\ref{eq1.7.0}). For $A=0.1<A_p$, one sees that the optimal path is localized around the initial position over a long-time window, and then escapes to the final position sharply at a late time. For $A>A_p$, the paths are degenerate, implying that there exist multiple optimal paths with the same $A$. In Fig.\ref{fig6}, we show three different optimal paths with different values of $\lambda$ but with the same $A=0.8$. The actions $S$ along these three paths can be obtained in terms of Eq.(\ref{eq4.4.2}), which are $S \approx 75.7$, $707.0$, and $1293.2$, and the corresponding $\lambda$ are approximately $115$, $389$, and $1294$, respectively. From the information, we can conclude that the path with the smallest $\lambda$ is dominant.

While it is non-Gaussian, one can expect that $P(A)$ around $A=A_p$ is approximately  Gaussian due to the central limit theorem. To the end, we expand $f_1(\lambda)$ in Eq.(\ref{eq4.4}) around $\lambda=0$ 
\begin{eqnarray}\label{eq4.5}
A(\lambda) = \frac{{13x_f^2{t_f}}}{{35}} + \frac{{59x_f^2t_f^5}}{{40425}}{\lambda}+o(\lambda^2).
\end{eqnarray} 
From Eq.(\ref{eq4.5}), we obtain 
\begin{eqnarray}\label{eq4.6}
\lambda  = \frac{{1155\left( {35A - 13x_f^2{t_f}} \right)}}{{59x_f^2t_f^5}}.
\end{eqnarray} 
Finally, we obtain the distribution of $\mathcal{A}=A$ by calculating the action in Eq.(\ref{eq1.2}),
\begin{eqnarray}\label{eq4.7}
-\ln P(A)  \simeq  \frac{3 x_f^2}{D t_f^3} + \frac{{40425{{\left( {A - {A_p}} \right)}^2}}}{{236Dt_f^5x_f^2}},
\end{eqnarray} 
where  $A_p$ is given in Eq.(\ref{eq4.4.1}).  Obviously, the Gaussian approximation holds only when $A$ is close to $A_p$. For $A$ away from $A_p$, $P(A)$ is no longer a Gaussian distribution, which is asymmetric with respect to $A_p$, as shown in Fig.\ref{fig5}(b). In Fig.\ref{fig7}(a), we show that the distribution of $A=\int_{0}^{t_f} x(t)^2 dt$, $P(A) \simeq e^{-S}$, for three different $D$, with $x_0=v_0=v_f=0$ and $x_f=t_f=1$ are fixed. The simulations results are also shown in Fig.\ref{fig7}(a), which are in excellent agreement with the analytical result.

We now consider the partially constrained case where the final velocity $v_f$ at $t=t_f$ is not fixed. For simplicity, we again set $x_0=v_0=0$. Four unknown coefficients can be determined by the boundary conditions: $x(0)=0$, $v(0)=0$, $x(t_f)=x_f$, and $\ddot x(t_f)=0$, which gives
\begin{eqnarray}\label{eq4.9}
\left\{ \begin{gathered}
{c_1} = \frac{{{x_f}\left[ {\sin \left( {\Lambda {t_f}} \right) + \sinh \left( {\Lambda {t_f}} \right)} \right]}}{{2\sinh \left( {\Lambda {t_f}} \right)\cos \left( {\Lambda {t_f}} \right) - 2\sin \left( {\Lambda {t_f}} \right)\cosh \left( {\Lambda {t_f}} \right)}} \hfill \\
{c_2} = \frac{{{x_f}\left[ {\cos \left( {\Lambda {t_f}} \right) + \cosh \left( {\Lambda {t_f}} \right)} \right]}}{{2\sin \left( {\Lambda {t_f}} \right)\cosh \left( {\Lambda {t_f}} \right) - 2\sinh \left( {\Lambda {t_f}} \right)\cos \left( {\Lambda {t_f}} \right)}} \hfill \\
{c_3} = \frac{{{x_f}\left[ {\cos \left( {\Lambda {t_f}} \right) + \cosh \left( {\Lambda {t_f}} \right) + \sin \left( {\Lambda {t_f}} \right) + \sinh \left( {\Lambda {t_f}} \right)} \right]}}{{4\sin \left( {\Lambda {t_f}} \right)\cosh \left( {\Lambda {t_f}} \right) - 4\sinh \left( {\Lambda {t_f}} \right)\cos \left( {\Lambda {t_f}} \right)}} \hfill \\
{c_4} = \frac{{{x_f}\left[ {1 + {e^{\Lambda {t_f}}}\cos \left( {\Lambda {t_f}} \right) - {e^{\Lambda {t_f}}}\sin \left( {\Lambda {t_f}} \right)} \right]}}{{2\cos \left( {\Lambda {t_f}} \right)\left( {{e^{2\Lambda {t_f}}} - 1} \right) - 2\sin \left( {\Lambda {t_f}} \right)\left( {{e^{2\Lambda {t_f}}} + 1} \right)}} \hfill \\ 
\end{gathered}  \right.
\end{eqnarray} 
for $\lambda>0$, and
\begin{eqnarray}\label{eq4.9.1}
\left\{ \begin{gathered}
{c_1} =  - {c_2} = \frac{{{x_f}\cos \left( {\Lambda {t_f}} \right)\cosh \left( {\Lambda {t_f}} \right)}}{{\sinh \left( {2\Lambda {t_f}} \right) - \sin \left( {2\Lambda {t_f}} \right)}} \hfill \\
{c_3} = \frac{{{x_f}\sin \left( {\Lambda {t_f}} \right)\left[ {1 + \cosh \left( {2\Lambda {t_f}} \right) - 2\cot \left( {\Lambda {t_f}} \right) + \sinh \left( {2\Lambda {t_f}} \right)} \right]}}{{2\left[ {\sinh \left( {2\Lambda {t_f}} \right) - \sin \left( {2\Lambda {t_f}} \right)} \right]\left[ {\cosh \left( {\Lambda {t_f}} \right) + \sinh \left( {\Lambda {t_f}} \right)} \right]}} \hfill \\
{c_4} = \frac{{{x_f}\left[ {\cosh \left( {\Lambda {t_f}} \right)\cos \left( {\Lambda {t_f}} \right) + \cosh \left( {\Lambda {t_f}} \right)\sin \left( {\Lambda {t_f}} \right) + \cos \left( {\Lambda {t_f}} \right)\sinh \left( {\Lambda {t_f}} \right)} \right]}}{{\sin \left( {2\Lambda {t_f}} \right) - \sinh \left( {2\Lambda {t_f}} \right)}} \hfill \\ 
\end{gathered}  \right. 
\end{eqnarray} 
for $\lambda<0$, where $\Lambda$ is defined in Eq.(\ref{eq4.1.1}).

Substituting the four coefficients in Eq.(\ref{eq4.9}) into Eq.(\ref{eq4.1}) and then performing the integral in Eq.(\ref{eq3.1}), we can establish the relation between $A$ and $\Lambda$, 
\begin{eqnarray}\label{eq4.10}
f_2(\tilde \Lambda) =\left\{ \begin{array}{lll} \frac{{x_f^2{t_f}{e^{2\tilde \Lambda }}}}{{2\tilde \Lambda {{\left[ {\sin \tilde \Lambda \left( {{e^{2\tilde \Lambda }} + 1} \right) - \cos \tilde \Lambda \left( {{e^{2\tilde \Lambda }} - 1} \right)} \right]}^2}}}\left\{ \tilde \Lambda \cos \left( {2\tilde \Lambda } \right) + 6\cosh \tilde \Lambda \sin \tilde \Lambda  + 3{{\cosh }^2}\tilde \Lambda \sin \left( {2\tilde \Lambda } \right) \right.  \\ \left. - \tilde \Lambda \left[ {\cosh \left( {2\tilde \Lambda } \right) + 4\sin \tilde \Lambda \sinh \tilde \Lambda } \right] - 3\cos \tilde \Lambda \left[ {2\sinh \tilde \Lambda  + \cos \tilde \Lambda \sinh \left( {2\tilde \Lambda } \right)} \right] \right\},    & \lambda>0,  \\
\frac{{x_f^2{t_f}}}{{16\tilde \Lambda {{\left[ {\sin \left( {2\tilde \Lambda } \right) - \sinh \left( {2\tilde \Lambda } \right)} \right]}^2}}}\left\{ 8\tilde \Lambda \cos \left( {2\tilde \Lambda } \right) - 2\cosh \left( {2\tilde \Lambda } \right)\left[ {4\tilde \Lambda  + 3\sin \left( {2\tilde \Lambda } \right)} \right] \right.  \\ \left. - 3\left[ {4\sin \left( {2\tilde \Lambda } \right) + \sin \left( {4\tilde \Lambda } \right)} \right] + 2\left[ {6 + 3\cos \left( {2\tilde \Lambda } \right) - 4\tilde \Lambda \sin \left( {2\tilde \Lambda } \right)\sinh \left( {2\tilde \Lambda } \right) + 3\sinh \left( {4\tilde \Lambda } \right)} \right] \right\},    & \lambda<0,  \\ 
\end{array}  \right. 
\end{eqnarray} 
where again $\tilde \Lambda= \Lambda t_f$. According to Eq.(\ref{eq1.2}), the action can be computed as
\begin{eqnarray}\label{eq4.11}
S(\tilde \Lambda) =\left\{ \begin{array}{lll} \frac{{x_f^2{\Lambda ^3}}}{{4D{{\left[ {\sin \tilde \Lambda \left( {{e^{2\tilde \Lambda }} + 1} \right) - \cos \tilde \Lambda \left( {{e^{2\tilde \Lambda }} - 1} \right)} \right]}^2}}}\left\{ \tilde \Lambda \cos \left( {2\tilde \Lambda } \right) - \tilde \Lambda \cosh \left( {2\tilde \Lambda } \right) - {{\cosh }^2}\tilde \Lambda \sin \left( {2\tilde \Lambda } \right) \right.  \\ \left. + 2\cos \tilde \Lambda \sinh \tilde \Lambda  - 2\sin \tilde \Lambda \left[ {\cosh \tilde \Lambda  + 2\tilde \Lambda \sinh \tilde \Lambda  + {{\cos }^2}\tilde \Lambda \sinh \left( {2\tilde \Lambda } \right)} \right] \right\},    & \lambda>0,  \\
\frac{{{e^{4\tilde \Lambda }}x_f^2{\Lambda ^3}}}{{4D{{\left[ {{e^{4\tilde \Lambda }} - 2{e^{2\tilde \Lambda }}\sin \left( {2\tilde \Lambda } \right) - 1} \right]}^2}}}\left\{ 8\tilde \Lambda \cosh \left( {2\tilde \Lambda } \right) - \sin \left( {4\tilde \Lambda } \right) + 4\sinh \left( {2\tilde \Lambda } \right) + 2\cos \left( {2\tilde \Lambda } \right)\left[ {\sinh \left( {2\tilde \Lambda } \right) - 4\tilde \Lambda } \right] \right.  \\ \left. - 2\sin \left( {2\tilde \Lambda } \right)\left[ {2 + \cosh \left( {2\tilde \Lambda } \right) - 4\tilde \Lambda \sinh \left( {2\tilde \Lambda } \right) + \sinh \left( {4\tilde \Lambda } \right)} \right] \right\},    & \lambda<0,  \\ 
\end{array}  \right. 
\end{eqnarray}

Taking the series expansion around $\lambda=0$ yields the Gaussian approximation, 
\begin{eqnarray}\label{eq5.0}
A = \frac{{33x_f^2{t_f}}}{{140}} + \frac{{1037x_f^2t_f^5}}{{970200}}{\lambda } + o\left( {{\lambda ^2}} \right),
\end{eqnarray} 
which results in 
\begin{eqnarray}\label{eq5.1}
\lambda  = \frac{{3465\left( {140A - 33x_f^2{t_f}} \right)}}{{1037x_f^2t_f^5}}.
\end{eqnarray}
Finally, we obtain the distribution of $\mathcal{A}=A$ by calculating the action in Eq.(\ref{eq1.2}),
\begin{eqnarray}\label{eq5.2}
-\ln P(A)  \simeq  \frac{3 x_f^2}{4D t_f^3}+ \frac{{121275{{\left( {A - {A_p}} \right)}^2}}}{{2074Dt_f^5x_f^2}},
\end{eqnarray} 
where 
\begin{eqnarray}\label{eq5.3}
A_p=\frac{{33}}{{140}}x_f^2{t_f}
\end{eqnarray} 
is the typical value of $A$ for the partially constrained case.

\begin{figure}
	\centerline{\includegraphics*[width=0.8\columnwidth]{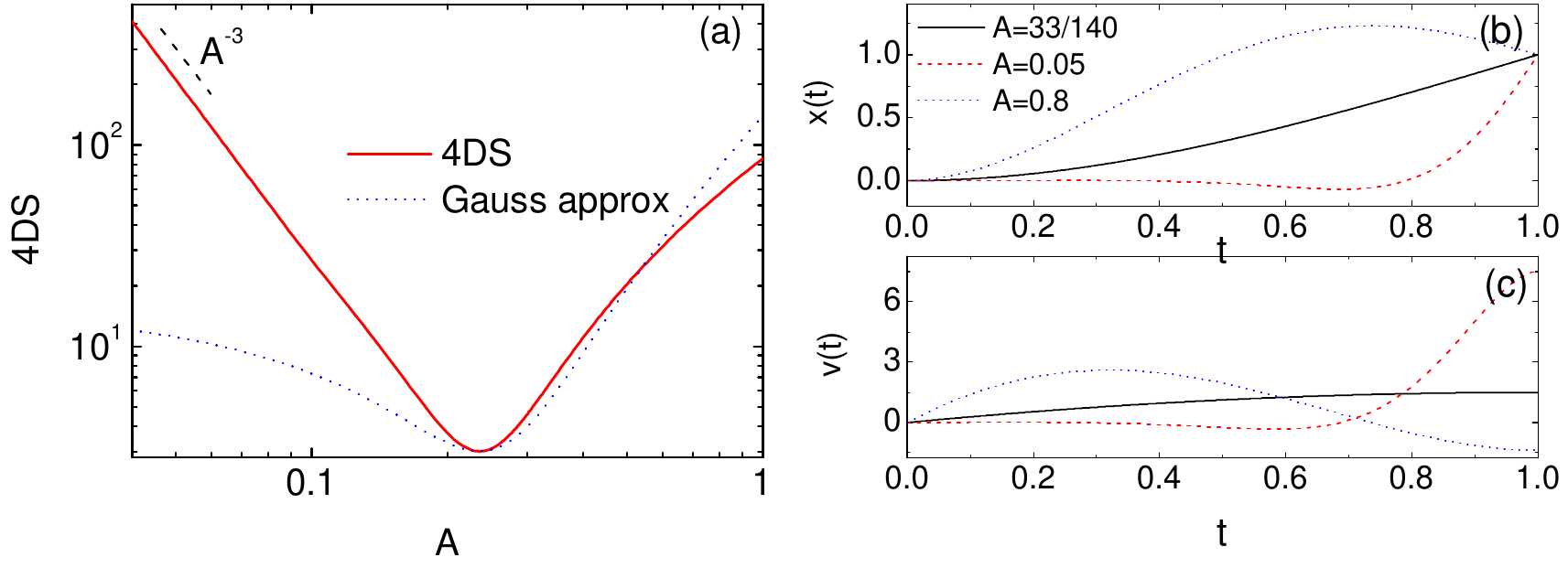}}
	\caption{Partial constraint with $n=2$. (a) The action as a function of $A$ (solid line). The dotted line gives the result of the Gaussian approximation. (b) The optimal path $x(t)$ and (c) the optimal velocity $v(t)$ for different values of $A$.  The other parameters are: $x_0=v_0=0$, $x_f=1$, and $t_f=1$.  \label{fig8}}
\end{figure}

\begin{figure}
	\centerline{\includegraphics*[width=0.6\columnwidth]{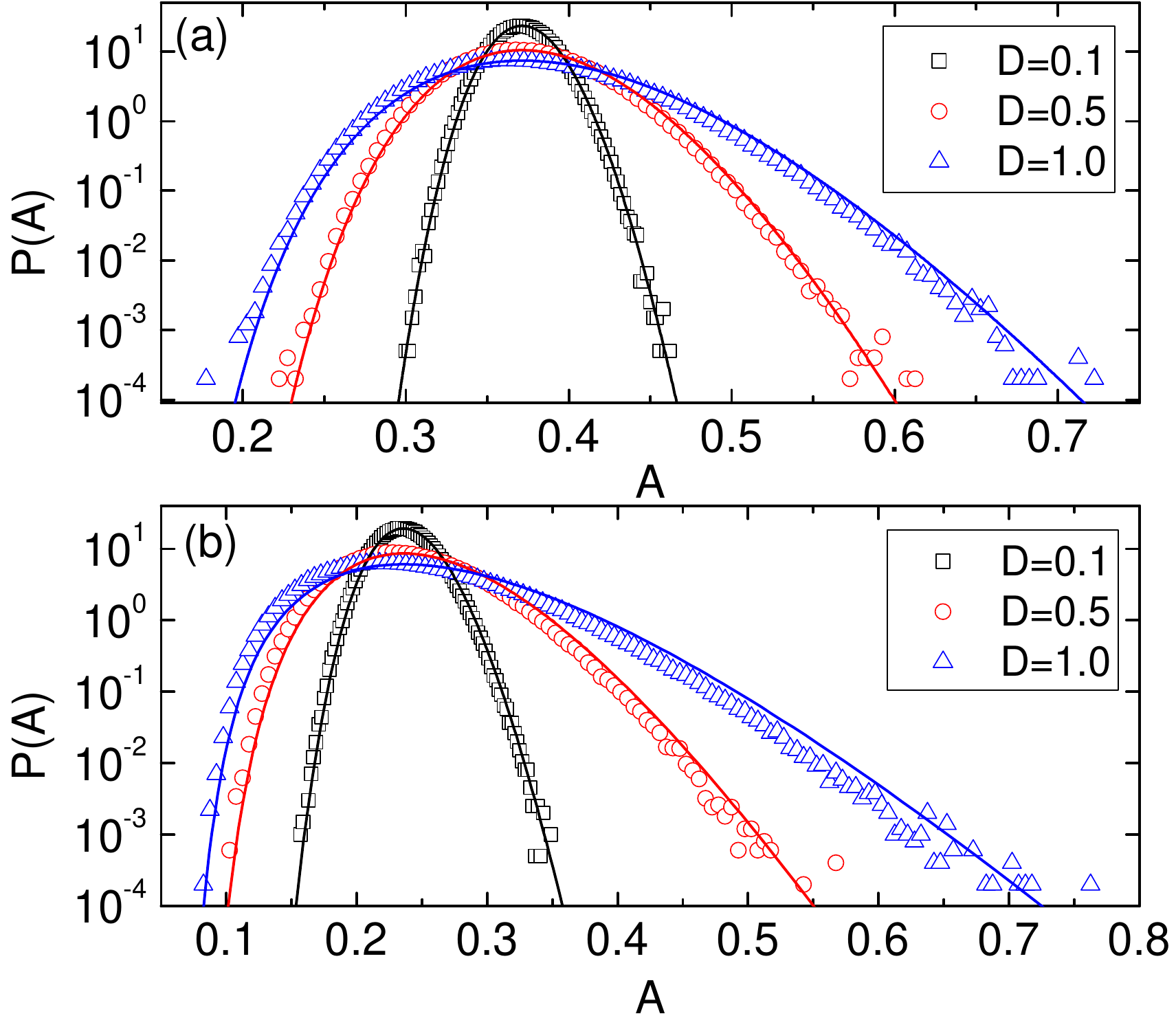}}
	\caption{Distribution of $A=\int_{0}^{t_f} x^2(t) dt$ for totally (a) and partially (b) constrained trajectories of random acceleration process for three different $D=0.1, 0.5, 1$. In (a), $x_0=v_0=0$, $x_f=1$, $v_f=0$, and $t_f=1$. In (b), $x_0=v_0=0$, $x_f=1$, $t_f=1$, and $v_f$ is a free parameter. Lines and symbols correspond to the theoretical and simulation results, respectively. \label{fig7}}
\end{figure}

The dependences of $A$ and $S$ on $\lambda$ are similar to the case of total constraint. In Fig.\ref{fig8}(a), we plot the action as a function of $A$. Again, for $A$ is away from its typical value $A_p$, the Gaussian approximation is no longer established. In the small-$A$ limit, the scaling relation in Eq.(\ref{eq4.8}) still exists. In Fig.\ref{fig8}(b), we show three typical optimal path $x^{*}(t)$ and the optimal velocity $v^{*}(t)$, corresponding to three different values of $A$: $A=A_p=33/140 \approx0.236$, 0.05, and 0.8, where $x_0=v_0=0$, $x_f=t_f=1$, and $v_f$ is a free parameter. In Fig.\ref{fig7}(b), we plot the distributions of $A$ with three different $D$ and compare them with simulations. The simulations agree well with the theoretical predictions.

\subsection{$n \geq 3$}
For $n \geq 3$, we are not able to solve analytically the Euler-Lagrange equation \ref{eq3.3}. Instead, we numerically obtain the solution via shooting method. The initial conditions $x(0)=x_0$ and $x'(0)=v(0)=v_0$ are known, but the other two initial conditions $x''(0)=b$ and $x'''(0)=c$ are unknown. Our goal is to determine $b$ and $c$
such that the other boundary conditions are fulfilled: i.e., $x(t_f)=x_f$ and $x'(t_f)=v(t_f)=v_f$ for the total constraint, and $x(t_f)=x_f$ and $x''(t_f)=a(t_f)=0$ for the partial constraint. In fact, this is a problem of finding roots for a set of appropriately defined functions. To the end, we define the functions $F_1=F_1(b,c)$ and $F_2=F_2(b,c)$ such that 
\begin{eqnarray}\label{eq6.1}
F_1(b,c)=x(t_f)-x_f
\end{eqnarray} 
and
\begin{eqnarray}\label{eq6.2}
F_2(b,c) =\left\{ \begin{array}{lll} x'(t_f)-v_f,    & {\rm{for}} \, {\rm{total}} \, {\rm{constraint}},  \\
x''(t_f),    & {\rm{for}} \, {\rm{partial}} \, {\rm{constraint}},  \\ 
\end{array}  \right. 
\end{eqnarray} 
where $x(t_f)$, $x'(t_f)$, and $x''(t_f)$ are functions of $b$ and $c$, obtained by numerically integrating the Euler-Lagrange equation \ref{eq3.3}, $x_f$ and $v_f$ are the prescribed boundary conditions at the ending time $t_f$. To determine the values of $b$ and $c$ that solves $F_1(b,c)=0$ and $F_2(b,c)=0$, we iterate using the Newton's method, given by  
\begin{eqnarray}\label{eq6.3}
\left( {\begin{array}{*{20}{c}}
	{{b_{k + 1}}} \\ 
	{{c_{k + 1}}} 
	\end{array}} \right) = \left( {\begin{array}{*{20}{c}}
	{{b_k}} \\ 
	{{c_k}} 
	\end{array}} \right) - {J^{ - 1}}\left( {{b_k},{c_k}} \right)\left( {\begin{array}{*{20}{c}}
	{{F_1}\left( {{b_k},{c_k}} \right)} \\ 
	{{F_2}\left( {{b_k},{c_k}} \right)} 
	\end{array}} \right),
\end{eqnarray} 
where the inverse of the Jacobian matrix $J^{-1}$ can be obtained by numerically evaluating the Jacobian 
\begin{eqnarray}\label{eq6.4}
J = \left( {\begin{array}{*{20}{c}}
	{\frac{{\partial {F_1}}}{{\partial b}}}&{\frac{{\partial {F_1}}}{{\partial c}}} \\ 
	{\frac{{\partial {F_2}}}{{\partial b}}}&{\frac{{\partial {F_2}}}{{\partial c}}} 
	\end{array}} \right). 
\end{eqnarray} 
Once the Euler-Lagrange equation \ref{eq3.3} are solved, we can calculate the functionals $A$ and the action $S$ from the equations (\ref{eq3.1}) and (\ref{eq1.2}), respectively. We have validated the numerical method by comparing numerical results with the analytical solutions for $n=1$ and 2.

\begin{figure}
	\centerline{\includegraphics*[width=0.6\columnwidth]{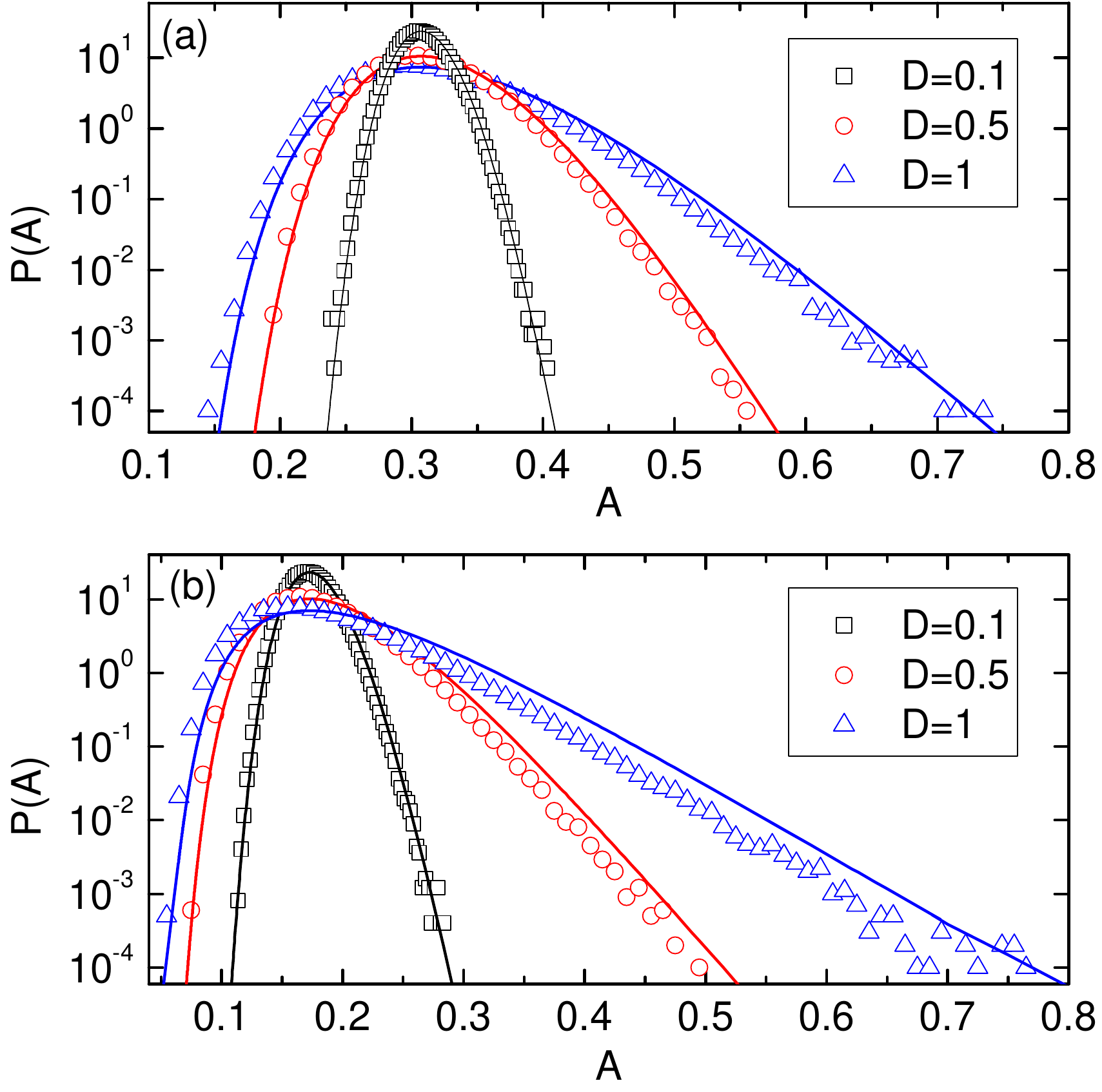}}
	\caption{Distribution of $A=\int_{0}^{t_f} x^3(t) dt$ for totally (a) and partially (b) constrained trajectories of random acceleration process for three different $D=0.1, 0.5, 1$. In (a), $x_0=v_0=0$, $x_f=1$, $v_f=0$, and $t_f=1$. In (b), $x_0=v_0=0$, $x_f=1$, $t_f=1$, and $v_f$ is a free parameter. Lines and symbols correspond to the theoretical and simulation results, respectively. \label{fig9}}
\end{figure}

In Fig.\ref{fig9} (a) and (b), we show the distribution of $A=\int_{0}^{t_f} x^3(t) dt$ for totally and partially constrained trajectories, respectively, with three different $D=0.1, 0.5, 1$. The lines correspond to the solution obtained by shooting method, and symbols to the simulation results. Clearly, the distribution for $n=3$ also exhibits the non-Gaussian feature. We also test the behavior of $-\ln P(A)$ in the limit of $A \to 0$, and find that the power-law divergence in Eq.(\ref{eq4.8}) holds as well.

\section{Conclusions}
In conclusion, we have employed the OFM to study the short-time deviations of the functionals $\mathcal{A}=\int_{0}^{t_f} x^n(t) dt$ along two classes of constrained trajectories $x(t)$ of RAP with a fixed duration $t_f$. Analytical calculations can be carried out for $n=1, 2$, and numerics for $n \geq 3$. For $n=1$, the functionals shows the usual Gaussian distribution. For $n \geq 2$,  the distribution $P(\mathcal{A}=A)$ is no longer a Gaussian function. The non-Gaussian feature is manifested by the asymmetry of the distribution function with respect to the most likely value of $A$. In the limit of $A \to 0$, $P(A)$ seems to be a universal essential singularity, $\ln P(A) \sim A^{-3}$. In addition to the distributions, the OFM enables us to predict optimal paths which provide valuable insights into the physics of RAP under specific constraints which
push the RAP into large-deviation regimes. Some interesting findings are the optimal paths in the large-deviation regimes. Especially for $n=2$, the optimal path for a given small-$A$ exhibits a localized pattern: the state of motion  remains almost unchanged for a long period of time at the beginning, followed by an abrupt change of the state of motion. When $A$ is much larger than its typical value, the optimal paths are 
degenerate, i.e., there exist multiple paths corresponding to the same value of $A$ but with different actions (or probability densities). We have identified that one of these degenerate optimal paths is dominant, and the others are exponentially unlikely.  

In the future, it would be very interesting to extend the OFM in two directions. On the one hand, since RAP in higher spatial dimensions is more relevant in real systems, such as motion of a living cell or a polymer, it is therefore desirable that the applications of the OFM in the high-dimensional RAP. On the other hand, a natural idea is to extend the RAP to higher-order processes \cite{bray2013persistence}, described by $x^{(n)}=\sqrt{2D} \xi(t)$ with $n>2$ where the superscript $(n)$ denotes the $n$th derivative with
respect to time. The development of OFM in the higher-order processes is also anticipated.

\appendix
\section{Euler-Lagrange equation}\label{appendix1}
Action functionals is defined as the integral of the Lagrangian along the path $x(t)$,
\begin{eqnarray}\label{eqa1.1}
S= \int_{0}^{t_f} \mathcal{L}(x,\ddot x)  dt,
\end{eqnarray} 
where we have assumed that the Lagrangian depends explicitly on $x$ and $\ddot x$. The variation of the action is written as
\begin{eqnarray}\label{eqa1.2}
\delta S= \int_{0}^{t_f} \left( \frac{{\partial \mathcal{L}}}{{\partial x}}\delta x + \frac{{\partial \mathcal{L}}}{{\partial \ddot x}}\delta \ddot x \right)   dt.
\end{eqnarray} 
Performing two integrations in parts in the second term, we obtain
\begin{eqnarray}\label{eqa1.3}
\delta S= \int_0^{t_f} {\left( {\frac{{\partial \mathcal{L}}}{{\partial x}} + \frac{{{d^2}}}{{d{t^2}}}\frac{{\partial \mathcal{L}}}{{\partial \ddot x}}} \right)\delta x\left( t \right)dt - \frac{{{d}}}{{d{t}}}\frac{{\partial \mathcal{L}}}{{\partial \ddot x}}\delta x} \left( t \right)\left| {_0^{t_f}} \right. + \frac{{\partial \mathcal{L}}}{{\partial \ddot x}}\left. {\delta \dot x\left( t \right)} \right|_0^{t_f}.
\end{eqnarray} 
Each of the three terms in the variation must vanish independently for arbitrary $\delta x$. The first term leads to the Euler-Lagrange equation
\begin{eqnarray}\label{eqa1.4}
{\frac{{\partial \mathcal{L}}}{{\partial x}} + \frac{{{d^2}}}{{d{t^2}}}\frac{{\partial \mathcal{L}}}{{\partial \ddot x}}}=0.
\end{eqnarray} 
The second term in Eq.(\ref{eqa1.3}) should vanish when the starting and final positions are fixed such that $\delta x(0)=\delta  x({t_f})=0$. For the total constraint, the starting and final velocities are also fixed such that $\delta \dot x(0)=\delta  \dot x({t_f})=0$, and thus the third term in Eq.(\ref{eqa1.3}) vanishes. For the partial constraint, the final velocity is not fixed, and the third term brings an extra boundary condition,
\begin{eqnarray}\label{eqa1.5}
{\left. {\frac{{\partial L}}{{\partial \ddot x}}} \right|_{t ={t_f}}} = 0.
\end{eqnarray}

\section{Effective Langevin equation for constrained stochastic processes}\label{appendix2}
Consider an $n$-dimensional Langevin equation, 
\begin{eqnarray}\label{eq.a1}
{d}{\mathbf{X}}(t)={\mathbf{u}} dt+ \sigma d{\mathbf{W}}(t)
\end{eqnarray}
where $\mathbf{X}(t) \in \mathbb{R}^n$ is the state variable of the system at time $t$,  $\mathbf{u} \in \mathbb{R}^n$ is the drift velocity, $\sigma\in \mathbb{R}^{n \times m}$ is the noise matrix, and $\mathbf{W}(t)\in \mathbb{R}^m$ are $m$ independent Wiener processes. For the constrained stochastic process conditioned on the initial state and final state, it is inefficient by directedly simulating the original Langevin \ref{eq.a1}. To overcome the difficulty, Majumdar and Orland proposed a novel method to 
generate the constrained paths by deriving the effective Langevin equation \cite{majumdar2015effective}. For completeness, in this appendix we will outline the main result in \cite{majumdar2015effective}, and at the same time we will extend to  more general situations where partial constraints is present in high-dimensional systems.

We now consider a constrained stochastic process that starts from $\mathbf{X}_0$ at $t=0$ and ends at $\mathbf{X}_f$ at $t=t_f$. Let us denote by $P_b(\mathbf{X},t)$ the probability density function of $\mathbf{X}$ at time $t$ with $0<t<t_f$. The constrained path can be decomposed into two parts: (i) the first segment
starts from $\mathbf{X}_0$ at time 0 and ends at $\mathbf{X}$ at time $t$, and the second segment starts from $\mathbf{X}$ at time $t$ and ends $\mathbf{X}_f$ at time $t_f$. Thus, $P_b(\mathbf{X},t)$ can be written as 
\begin{eqnarray}\label{eq.a2}
P_b(\mathbf{X},t)=\frac{P(\mathbf{X},t|\mathbf{X}_0,0) P(\mathbf{X}_f,t_f|\mathbf{X},t)}{P(\mathbf{X}_f,t_f|\mathbf{X}_0,0)},
\end{eqnarray}
where $P(\mathbf{X},t|\mathbf{X}_0,0)$ the probability density function of $\mathbf{X}$ at time $t$ for unconstrained processes, providing the system has started with $X_0$ at $t=0$.
For simplifying notation, we define $P(\mathbf{X},t|)=P(\mathbf{X},t|\mathbf{X}_0,0)$ and $Q(\mathbf{X},t)=P(\mathbf{X}_f,t_f|\mathbf{X},t)$.
$P(\mathbf{X},t)$ and $Q(\mathbf{X},t)$ satisfy the forward and backward Fokker-Planck equations, respectively, given by \cite{gardiner1985handbook}
\begin{eqnarray}\label{eq.a3}
\begin{gathered}
\frac{{\partial P\left( {\mathbf{X},t} \right)}}{{\partial t}} =  - \sum\limits_i {\frac{{\partial \left( {{u _i}P} \right)}}{{\partial {x_i}}}}  + \frac{1}{2}\sum\limits_i {\sum\limits_j {\frac{{{\partial ^2}\left( {{B_{ij}}P} \right)}}{{\partial {x_i}\partial {x_j}}}} },  \hfill \\
\frac{{\partial Q\left( {\mathbf{X},t} \right)}}{{\partial t}} =  - \sum\limits_i {{\mu _i}\frac{{\partial Q}}{{\partial {x_i}}}}  + \frac{1}{2}\sum\limits_i {\sum\limits_j {{B_{ij}}\frac{{{\partial ^2}Q}}{{\partial {x_i}\partial {x_j}}}} } , \hfill \\ 
\end{gathered} 
\end{eqnarray}
where $B=\sigma \sigma^{\top}$ is the diffusion matrix. Substituting Eq.(\ref{eq.a3}) into  Eq.(\ref{eq.a2}), we can obtain the corresponding Fokker-Planck equation for $P_b(\mathbf{X},t)$, 
\begin{eqnarray}\label{eq.a4}
\frac{{\partial {P_b}\left( {\mathbf{X},t} \right)}}{{\partial t}} =  - \sum\limits_i {\frac{\partial }{{\partial {x_i}}}\left[ {{u _i} + \sum\limits_j {{B_{ij}}\frac{{\partial \ln Q}}{{\partial {x_j}}}} } \right]{P_b} } +  \frac{1}{2}\sum\limits_i {\sum\limits_j {\frac{{{\partial ^2}\left( {{B_{ij}}{P_b}} \right)}}{{\partial {x_i}\partial {x_j}}}} } .
\end{eqnarray}
The effective Langevin equation corresponding to Eq.(\ref{eq.a3}) reads
\begin{eqnarray}\label{eq.a5}
d {\mathbf{X}_b}(t) = \tilde {\mathbf{u}} dt  + \sigma d\mathbf{W}(t),
\end{eqnarray}
where
\begin{eqnarray}\label{eq.a6}
{{\tilde u }_i} = {u _i} + \sum\limits_j {{B_{ij}}\frac{{\partial \ln Q}}{{\partial {x_j}}}} 
\end{eqnarray}
is the $i$th component of the effective drift velocity.

We now relax the constraints in the final state of the system. To the end, we assume a set of components of $\mathbf{X}_f$ is fixed, saying $\mathbf{X}_f^c$, and the other components of $\mathbf{X}_f$ is free, saying $\mathbf{X}_f^{nc}$, such that $\mathbf{X}_f=\mathbf{X}_f^c  \bigoplus \mathbf{X}_f^{nc}$. The probability density function of $\mathbf{X}$ at time $t$ for the constrained stochastic path is written as
\begin{eqnarray}\label{eq.a8}
P_b(\mathbf{X},t)=\frac{P(\mathbf{X},t|\mathbf{X}_0,0) \int d{\mathbf{X}_f^{nc}} P(\mathbf{X}_f,t_f|\mathbf{X},t)}{\int d{\mathbf{X}_f^{nc}} P(\mathbf{X}_f,t_f|\mathbf{X}_0,0)}.
\end{eqnarray}
Using the same method as before, we can also obtain the Fokker-Planck equation for $P_b(\mathbf{X},t)$ and the corresponding Langevin equation for the partially constrained case. The results are the same as Eq.(\ref{eq.a4}) and Eq.(\ref{eq.a5}) except that $Q(\mathbf{X},t)$ is now defined as $Q(\mathbf{X},t)=\int d{\mathbf{X}_f^{nc}}P(\mathbf{X}_f,t_f|\mathbf{X},t)$.

\begin{acknowledgments}     
This work was supported by the National Natural Science Foundation of China
(11875069), the Key Scientific Research Fund of Anhui Provincial Education
Department (2023AH050116), and Anhui Project (Grant No. 2022AH020009)
\end{acknowledgments}


\end{document}